\documentclass[a4paper,fleqn,usenatbib]{mnras}

\pdfoutput=1

\usepackage[T1]{fontenc}
\usepackage{ae,aecompl}

\usepackage{pdflscape} 

\usepackage{graphicx}	
\usepackage{amsmath}	
\usepackage{amssymb}	

\usepackage{mathrsfs}
\usepackage[english]{babel}

\usepackage[usenames]{color}

\newcommand{\f}[2]{\frac{#1}{#2}} 
\newcommand{\Ang}{\buildrel _{\circ} \over {\mathrm{A}}}

\title[Synthetic absorption lines of warm absorbers]
{Synthetic absorption lines for a clumpy medium:\\ 
 a spectral signature for cloud acceleration in AGN?}

\author[Waters, Proga, Dannen, \& Kallman]{
Tim Waters$^1$,\thanks{E-mail: waterst3@unlv.nevada.edu}
Daniel Proga$^1$,
Randall Dannen$^1$, \&
Timothy R. Kallman$^2$
\\
$^1$Department of Physics \& Astronomy, University of Nevada, Las Vegas, 
  4505 S. Maryland Pkwy, Las Vegas, NV, 89154-4002, USA\\
$^2$NASA Goddard Space Flight Center, Code 662, Greenbelt, MD 20771, USA
}

\date{Accepted XXX. Received YYY; in original form ZZZ}

\pubyear{2016}

\begin{document}
\label{firstpage}
\pagerange{\pageref{firstpage}--\pageref{lastpage}}
\maketitle

\begin{abstract}
There is increasing evidence that the highly ionized multiphase components of AGN 
disc winds may be due to thermal instability.  The ions responsible for forming the observed 
X-ray absorption lines may only exist in relatively cool clumps 
that can be identified with the so-called `warm absorbers'.  
Here we calculate synthetic absorption lines for such warm absorbers from first principles 
by combining 2D hydrodynamic solutions of a two-phase medium with a dense grid of 
photoionization models to determine the detailed ionization structure of the gas.
Our calculations reveal that cloud disruption, which leads to a highly complicated velocity 
field (i.e. a clumpy flow), will only mildly affect line shapes and strengths when the warm gas 
becomes highly mixed but not depleted.  Prior to complete disruption, 
clouds which are optically thin to the driving UV resonance lines 
will cause absorption at an increasingly blueshifted line of sight velocity as they are accelerated.
This behavior will imprint an identifiable signature on the line profile if warm absorbers are 
enshrouded in an even broader absorption line produced by a high column of intercloud gas.
Interestingly, we show that it is possible to develop a spectral diagnostic for cloud acceleration 
by differencing the absorption components of a doublet line,  
a result which can be qualitatively understood using a simple partial covering model.  
Our calculations also permit us to comment on 
the spectral differences between cloud disruption and ionization changes driven by flux variability.  
Notably, cloud disruption offers another possibility for explaining absorption line variability. 
\end{abstract}


\section{Introduction}
Many spectral properties of active galactic nuclei (AGN) are often explained by appealing 
to models involving high velocity clouds that can coexist with more highly ionized surrounding gas 
(for an overview see Krolik 1999 and Netzer 2013).  However, the AGN environment is very hostile 
to such clouds because the strong radiation field and the presence of velocity shear leaves them 
prone to various hydrodynamical instabilities (e.g., Krolik 1977, 1979, 1988; 
Mathews \& Blumenthal 1977; Mathews 1986).  Therefore, for the clouds to be present, 
they either need to be confined or efficiently regenerated.

In a few recent papers we have addressed the issue of could formation
and regeneration.  In particular, in Proga et al. (2008) we presented results from global
simulations of accretion flows irradiated by the AGN.  We found these flows can turn
into outflows due to radiation pressure. We also found that these outflows are clumpy
due to thermal instability (TI; Field 1965) and the radiation force. In our subsequent papers
we investigated the physics of these clouds
in more detail using global simulations (Kurosawa \& Proga 2008a; Barai et
al. 2012; Moscibrodzka 2013) as well as local simulations (Proga et al.
2014; Proga \& Waters 2015, hereafter PW15; Waters \& Proga 2016, hereafter WP16).
One of the main results of these studies is that TI and
radiation pressure can facilitate the regeneration and acceleration of clouds.

In PW15 we showed how to self-consistently account for the increase in radiation force 
(due primarily to resonance line scattering) as a cloud forms and ion opacity increases.  
In so doing, we verified very early studies (e.g., Krolik 1977, 1979; Mathews \& Blumenthal 1977) 
showing that radiatively accelerated planar clouds (i.e. `slabs') are unstable to the Rayleigh-Taylor 
instability on the leading edge and will thus be disrupted.  In WP16, we explored the effects of flux 
variability on the acceleration of clouds.  We found that, independent of the presence of a variable 
flux, the gas remains two-phase and becomes highly mixed. The complicated velocity field 
supplies perturbations that permit the continual production of new clouds via TI, offsetting any 
losses to evaporation back into the hot medium.  The long term state of the gas is a clumpy flow 
that accelerates at a nearly constant rate determined by the amplitude and frequency of the 
variations in the ionizing flux. 

In this paper, we set out to calculate synthetic absorption line profiles for the simulations of 
PW15 and WP16, which probe a temperature regime appropriate for modeling warm 
absorbers.  In particular, we sought to answer the following questions:
(1) is it possible to differentiate, spectroscopically, between an ideal, comoving cloud 
(i.e. of planar or round geometry),
a cloud being disrupted as it accelerates through its surrounding medium,
and a completely chaotic, clumpy flow?; 
(2) can absorption line variability due to cloud disruption be distinguished from that due to 
variability in the ionizing flux?
(3) can a standard partial covering analysis of our synthetic line profiles adequately 
reproduce the properties (namely, the velocity-dependent optical depth and 
covering fraction profiles) of our solutions?

As for the first question, our expectation is that cloud acceleration can in principle leave 
an absorption signature as the cloud is being disrupted, provided the absorption line is 
broader than the thermal width of the warm gas alone.   Namely, 
the clumps of gas with the highest opacity end up the fastest moving, 
so that if absorption from warm gas is overall enshrouded 
in a broader absorption line formed in the hot inter-cloud gas, there will be systematically 
more absorption on the blue side of line center.  Hence, the expected signature for the 
acceleration of a newly formed cloud is an initially symmetric line profile that becomes 
asymmetric with time as a somewhat deeper absorption feature (that tracks the warm cloud material) 
becomes increasingly blueshifted. 

To confirm this expectation and answer the other questions posed, it was necessary to 
perform photoionization calculations in order to calculate line opacities for the most abundant 
ions present at the temperature range of our simulations.  These calculations are described 
in \S{2}, along with our methods for calculating simulated absorption lines.  Our results are 
presented in \S{3}, followed by our discussion in \S{4} and conclusions in \S{5}.
 
\section{Methods}
Time-dependent, 2D hydrodynamical simulations of a two phase medium form the basis for 
our calculations of absorption line profiles from first principles, as they provide the density, 
velocity, and temperature fields of the gas.  This information is used to determine the opacity 
(and hence optical depth) of a specific ion by also running a dense grid of photoionization 
models using \textsc{xstar} (Kallman \& Bautista 2001) for an assumed set of atomic 
abundances and spectral energy distribution (SED).  The absorption line profiles are calculated 
by accounting for the attenuation of a uniform background source.  Below we describe these 
methods in detail, emphasizing the underlying physical assumptions made when applying 
photoionization modeling to dynamical flow solutions.  Finally, we discuss our procedure for 
making a comparison between our synthetic line profiles and a commonly used partial 
covering model (e.g., Barlow \& Sargent 1997; Hamann et al. 1997).  

\subsection{Hydrodynamical simulations}
\label{sec:sims}
Using \textsc{Athena} (Stone et al. 2008), we perform local cloud simulations similar to those 
presented in PW15 and WP16.  These calculations account for many of the relevant 
microphysical processes believed to be important in the environment of AGN: heating and 
cooling due to the dominant radiative processes 
(photoionization and recombination, line cooling, bremsstrahlung, and 
compton/inverse compton processes using the formula from Blondin (1994)), 
thermal conductivity between the cloud and the inter-cloud medium, and a radiation 
force self-consistently determined from the heating processes and the number of lines 
present at a given temperature.  Specifically, the radiative driving term was specified using 
the modified Castor, Abbot, \& Klein (1975) method of Owocki et al. (1988), 
with a reduction of the force multiplier due to strong photoionization as
computed by Stevens \& Kallman (1990) and Kallman (2006, private
communication), who also included a more realistic distribution of
driving lines.

We focus our analysis on two different initial configurations leading to a clumpy medium: 
(i) a round cloud and 
(ii) a `slab' configuration, which can be viewed as a local portion of a spherically symmetric 
shell of cloud material.  
Both configurations are arrived at by evolving a small isobaric perturbation in a homogenous, 
periodic box of nearly fully ionized gas that is thermally unstable.  

For the purpose of this paper we run new simulations that are basically reruns of those 
presented in PW15.  The main differences are that we use a square domain and for 
case (i) we take the amplitudes and growth rates of the initial perturbations to be identical in 
both the $x$ and $y$ directions to promote the formation of a round cloud. 

For each configuration, we also carry out an additional simulation that uses a time-varying 
ionizing flux, $F_{\rm{ion}}(t)$, in order to study the effects of variability.  This time dependence 
carries over to the photoionization parameter, $\xi = 4\pi F_{\rm{ion}}/n$, where $n$ is the gas 
number density.  As in WP16, we model the variability using a simple sinusoidal flux,
\begin{equation}
F_{\rm{ion}}(t) =  F_X(1 + A_X \sin(2\pi t/t_X)),
\label{eq:Fion}
\end{equation}
where $A_X$ and $t_X$ are the amplitude and period of flux oscillations about a mean flux 
$F_X = n_{eq} \xi_{eq}/4\pi \approx 7.82\times 10^8~\rm{erg~s^{-1}~cm^{-2}}$.  The subscript 
`eq' denotes values used to define the initial equilibrium state of the gas.  As described by PW15, 
$\xi_{eq} = 190$ is chosen based on numerical requirements to adequately resolve the interfaces 
between warm and hot gas.  The relation $n_{eq} T_{eq} = 10^{13}~\rm{cm^{-3}~K}$ typical of 
AGN (e.g., Krolik 1999) sets $n_{eq}$ because $T_{eq} = 1.93\times 10^5~\rm{K}$ is fixed by the 
radiative equilibrium curve once $\xi$ is chosen (see PW15).  We again adopt $A_X = 0.2$ and 
$t_X =5\,t_{th} \approx 3.5~\rm{days}$, where $t_{\rm{th}}$ is the thermal time, defined as the ratio 
of the internal energy density of the gas to the net volumetric heating rate on the radiative 
equilibrium curve at $(\xi_{eq},T_{eq})$. 

These simulations explored a parameter regime chosen such that the majority of lines contributing to 
the line driving are optically thin.  The gas is therefore too highly ionized to host an abundance of ions 
responsible for forming broad absorption lines (BALs) commonly observed in the UV 
(e.g., $\ion{S}{iv}$, $\ion{C}{iv}$, $\ion{O}{vi}$).  The strongest line, as determined by our 
photoionization calculations described below, is the X-ray resonance line 
$\ion{O}{viii}~\rm{Ly\alpha}$, which is a doublet with transitions at 
$18.9671\,\Ang$ and $18.9725\,\Ang$.  Our results are therefore confined to calculations for this line, 
commonly observed in warm absorbers (e.g., Turner \& Miller 2009).  

Warm absorbers are observed to have outflow velocities exceeding $100~\rm{km~s^{-1}}$ 
(e.g., Crenshaw et al. 2003).  Our cloud simulations are local, meaning they are performed in the 
comoving frame of the cloud and therefore apply to warm absorbers formed at any velocity, 
so long as it is not relativistically large.  We calculate line profiles as seen by a distant observer, 
and at time zero before the cloud is formed, we center the line profiles at zero velocity 
without loss of generality.

\subsection{Photoionization calculations}
\label{sec:Xstar}
Determining the detailed photoionization structure of our clumpy medium requires solving for, 
at every location in our domain, the fractional ion abundances and steady state level populations 
for a given set of elemental abundances.  For this, we employ version 2.35 of the public code 
\textsc{xstar} (Kallman \& Bautista 2001).  \textsc{xstar} was run in constant density mode with 
$n_{\rm{xstar} }= 10^{10}~\rm{cm}^{-3}$.  To be consistent with the heating and cooling rates 
adopted in our cloud simulations, we used a 10~keV Bremsstrahlung SED.  We assigned solar 
elemental abundances from Grevesse, Noels \& Sauval (1996).  

\textsc{xstar} calculates photoionization equilibrium, accounting for over 200,000 lines 
(Bautista \& Kallman 2001), for gas with a given photoionization parameter and temperature.  
In our simulation domain, every grid cell is characterized by the two parameters $(\xi,T)$ local 
to this cell.  We therefore ran a grid of models in $(\xi,T)$-space covering the full parameter 
space of our solutions.  This grid spanned 200 points logarithmically spaced over the range 
$\log \xi = 0$ to $\log \xi = 8$, and 94 points logarithmically spaced from 
$T = 5\times10^3~\rm{K}$ to $10^8~\rm{K}$.  
Since \textsc{Athena} has already determined the temperature of the gas,
\textsc{xstar} is run with the option set to keep the input temperature fixed.
\textsc{xstar} provides, in addition to the ionization structure, 
quantities such as the opacity, emissivity, excitation and dexcitation rates, and heating/cooling 
rates due to individual processes. 

The relevant value we extract from \textsc{xstar} is the comoving frame line center opacity
of a resonance line (uncorrected for stimulated emission), 
 \begin{align}
 \begin{split}
 \kappa_{\nu_0} &  = \f{\pi e^2}{m_e c} \f{n_1}{\rho}f_{12} \, \phi(\nu_0) \\
 & = \f{1}{\sqrt{\pi}}\left( \f{\pi e^2}{m_e c} \f{A\, n_{\rm{xstar}} }{\nu_0 (v_{th}/c) } f_{12}\,  \eta_{ion} \right)   .
\f{1}{\mu m_p n_{\rm{xstar}}}.
\end{split}
\label{eq:alpha_nu0}
\end{align}
Here, $\rho = \mu m_p n$ is the hydrodynamic density (we set $\mu = 1$ in this work), 
$f_{12}$ is the oscillator strength of the lower level of the line, 
$n_1$ is the level population of the ion's ground state, and $\phi(\nu)$ is the line profile 
that will be calculated separately (see \S{2.3}).  The level population is related to the number density 
from our hydrodynamical solutions through $n_1 = A\,\eta_{\rm{ion}}\,n$, where $A$ is the elemental 
abundance relative to hydrogen and $\eta_{\rm{ion}}$ is the fractional ion abundance.  
In the second line,
$v_{th} = \sqrt{2\,k\,T/m_{\rm{i}}}$ is the mean thermal velocity of an ion with atomic weight $m_{\rm{i}}$.
The term in parenthesis is the attenuation coefficient we extract from \textsc{xstar}.  
In the remaining term, we correct for the constant density assigned to \textsc{xstar}, $n_{\rm{xstar}}$.  
By applying the density correction in this way, we assume that the opacity is rather insensitive to $n$, 
depending instead on the ratio of the density of photons to the gas number density, i.e. the 
photoionization parameter.  Given $\xi$ and $T$, we determine the line center opacity from \textsc{xstar} 
output using a lookup table that bilinearly interpolates between the values in our grid.

Most photoionization modeling efforts explicitly assume a static slab configuration, 
as hydrodynamical effects such as compressional heating and conductive heat fluxes that 
characterize a dynamical flow are not accounted for by $\textsc{xstar}$.   
However, it is valid to account for the latter effects independently of $\textsc{xstar}$ using
time-dependent hydrodynamical simulations provided that photoionization equilibrium is maintained as the flow evolves.  
That is, our approach is justified when the relevant hydrodynamical time-scales  
are long compared to timescales governing all of the atomic processes.  

\begin{table}
 \caption{Characteristic recombination rates and comparison with dynamical timescales 
 for three representative \textsc{xstar} models.}
 \label{tab:example}
 \begin{tabular}{llcc}
  \hline
  Gas regime &Atom/ & $n_e \alpha$  & $t_{\rm{dyn}} / t_{\rm{rec}}$ \\
  ($T$ [K], $\xi$ [ergs cm $s^{-1}$]) & Ion & $[s^{-1}]$& \\
  \hline
  \hline
  Warm & \ion{H}{i} & $1.04 \times 10^{-3}$ & $1.1\times10^2$\\
  $(7.53\times10^4, 77.0)$ & \ion{O}{viii} & $1.21 \times 10^{-1}$ & $1.2\times10^4$\\
  \hline
  Evaporating & \ion{H}{i} & $4.45 \times 10^{-4}$ & 5.0\\
    $(1.93\times10^5, 190.0)$ & \ion{O}{viii} & $7.64 \times 10^{-2}$ & $8.6\times10^2$\\
  \hline
  Hot & \ion{H}{i} & $1.87 \times 10^{-4}$ & 8.6\\
  $(4.77\times10^5, 475.0)$  & \ion{O}{viii} & $4.56\times 10^{-2}$ & $2.1\times10^3$\\
  \hline

 \end{tabular}\\
\end{table}

The shortest timescale on which clouds can undergo structural changes is the sound crossing time 
across a layer of cloud material.  The thinnest cloud layers in our simulations belong to clouds about 
to undergo evaporation, which occurs on a longer timescale; the length scale for evaporation therefore 
determines the smallest cloud sound crossing time.  This length scale is the Field length, 
$\lambda_F = 2\pi \sqrt{\kappa T/\rho \Gamma}$ (Begelman \& McKee 1990), giving 
$t_{\rm{dyn}} = \lambda_F/c_s = 2\pi\sqrt{\kappa/(\gamma n k \Gamma)} \approx 10^4\,n_8^{-1/2}\, \rm{s}$, 
where $n_8$ is the number density in units of $10^8~\rm{cm}^{-3}$ and we have plugged in fiducial values for the 
thermal conductivity $\kappa$ and the volumetric heating rate $\Gamma$ on our radiative equilibrium curve at 
$\xi=190$ (see PW15).  Note that this time-scale is still sensitive to temperature since both $\kappa$ and 
$\Gamma$ have implicit temperature dependence.  We must compare $t_{\rm{dyn}}$ with the recombination 
time-scale, $t_{\rm{rec}}$, as it is the microphysical process that typically has the longest time-scale (e.g., Ferland 1979).  

Recombination rates per ion, $n_e \alpha$ (where $n_e$ is the free electron number density), 
as well as the ratios $t_{\rm{dyn}}/t_{\rm{rec}}$ for both hydrogen and $\ion{O}{viii}$ are tabulated 
for three different temperatures and corresponding photoionization parameters in Table~1.  
The middle row was calculated for $(T_{eq},\xi_{eq})$, the equilibrium values used to set the initial 
conditions of the simulations, and these values are characteristic of those for the evaporating gas.  
The top and bottom rows quote values for $T$ and $\xi$ typical of the warm and hot gas in the domain 
at any given time as the gas evolves.  For this gas, the dynamical time is defined as simply the domain 
sound crossing time, $L_x/c_s$ (at the appropriate sound speed).  We see that the necessary condition 
$t_{\rm{dyn}} > t_{\rm{rec}} \equiv (n_e \alpha)^{-1}$ is always satisfied by at least a factor of 5 for hydrogen.  
Thus, the validity of applying photoionization calculations to calculate the ionization structure of our 
dynamical flow solutions has been established, even for the most transient, evaporating gas in our simulations.   

\subsection{Absorption line profiles}
A formal solution of the radiative transfer equation that holds when the source function for local 
emission is zero reads
\begin{equation}
\mathscr{L}_\nu=\int_0^{L_y}\int_0^{L_x} \mathcal{I}_{\nu}(x,y)e^{-\tau_\nu(x,y)}dx\,dy,
\label{eq:Lnu_xy}
\end{equation}
where $\mathcal{I}_{\nu}(x,y)$ is the monochromatic specific intensity of the
background source.  
This integral is taken over an imaginary plane of the sky (located beyond the absorber) of linear 
dimensions $(L_x,L_y)$ as viewed by a distant observer, while evaluating the
optical depth $\tau_\nu(x,y) = \int \kappa_\nu \rho \, dz$ requires an integration of the opacity 
$\kappa_\nu(x,y,z)$ over the line of sight (LOS) coordinate $z$.   
Note that by neglecting the source function, we do not account for scattering and remission of the
line photons.  Therefore, we do not include certain effects such as the net expansion of the medium, 
which can allow photons to be first absorbed in the blue component, and then reemitted and absorbed 
by the red component in a different region of the flow (e.g., Castor and Lamers 1979).

The calculation of absorption lines simplifies when there is symmetry along the $y$-direction of the 
observer plane, the case for 2D simulations.  Integrating over $y$ and measuring $x$ in units of 
$L_x = 3.1\times10^{11}~\rm{cm}$ so that $x$ varies from 0 to 1 gives
\begin{equation}
\mathscr{L}_\nu=L_x\,L_y\int_0^1 \mathcal{I}_{\nu}(x)e^{-\tau_\nu(x)}dx,
\label{eq:Lnu_x}
\end{equation}
where now the $x$-coordinate defines a single LOS in our calculations.   
We assume that the background intensity is constant.  
Thus, by defining the normalized absorption line profile $I \equiv \mathscr{L}_\nu / (L_x\,L_y\,\mathcal{I}_{\nu})$, 
we arrive at 
\begin{equation}
I(\nu) = \int_0^1 e^{-\tau_\nu(x)}dx.
\label{eq:I_x}
\end{equation}

Evaluating the optical depth requires calculating 
the opacity $\kappa_\nu(x,z) = \kappa_{\nu_0}(x,z) \phi(\nu)/\phi(\nu_0)$ at every location in the domain.  
As described in \S{\ref{sec:Xstar}}, we directly determine $\kappa_{\nu_0}$ from \textsc{xstar}.  
The line profile function $\phi(\nu)$ is evaluated separately using our hydrodynamical variables.  
Every finite volume cell in our grid-based solution is treated as a parcel of gas that is thermally broadened 
according to its cell-centered temperature (between about $2\times10^5\rm{K}$ and $10^6\rm{K}$) 
and Doppler shifted away from line center according to its LOS velocity.  
Each local line profile is then specified as a Gaussian,
\begin{equation}
\phi(\nu) = \f{1}{\sqrt{\pi}} \f{1}{\Delta \nu_0} \exp\left(\f{-(\nu-\nu_D)^2}{\Delta \nu_0^2} \right),
\label{eqn:phi}
\end{equation}
where $\Delta \nu_0 = \nu_0 v_{th}/c$ is the thermal linewidth and $\nu_D =\nu_0(1 + v_z/c)$ is the Doppler 
shifted frequency at line center.  
We note that the width of the resulting net line profile is primarily set by the thermal velocity of the hot gas, 
$v_{th,hot} \approx 20-25~\rm{km~s^{-1}}$, but it can also be increased by the $z$-component 
of the local velocity field.

We use the trapezoidal rule to evaluate both the optical depth integral and the integral in equation \eqref{eq:I_x} 
numerically.  These integrations are performed at 150 different frequencies centered around $\nu_D$.  
In practice, we work in velocity space, binning the range $[v_c - 50~\rm{km~s^{-1}}, v_c + 50~\rm{km~s^{-1}}]$, 
where $v_c$ is the `cloud-tracking' velocity, or the velocity of the comoving frame that is used by \textsc{Athena} 
to keep the cloud centered on the domain.

\subsection{Doublet lines and the PPC model}
As mentioned, we apply the above methods to simulate absorption from $\ion{O}{viii}~\rm{Ly\alpha}$, 
a common X-ray doublet line, whose synthetic line profiles will be denoted $I_r$ and $I_b$.  
The subscripts `r' and `b' distinguish these doublet components, with $I_b$ having a slightly higher 
(bluer) frequency.  Many doublet lines have the property that the stronger line has an oscillator strength 
almost exactly twice that of the weaker line (due to relativistic effects, the fractional difference is 
$\sim10^{-4}$), and so the ratio of line-center opacities is also almost 2. 

Doublet lines produced in a clumpy medium are especially interesting, as revealed by the so-called 
pure partial covering (PPC) model (Barlow \& Sargent 1997; Hamann et al. 1997; de Kool et al. 2002), 
a widely used model for estimating the ionic column densities of AGN outflows (e.g., Crenshaw et al. 2003).  
According to this model,
the absorption line profiles for the two transitions are 
 \begin{align}
 \begin{split}
& I_r = (1-C_\nu)  + C_\nu\,e^{-\tau_{\nu,r}};\\
& I_b = (1-C_\nu) + C_\nu\,e^{-2\,\tau_{\nu,r}}.
\end{split}
\label{eq:ppc_model}
\end{align} 
These expressions follow from equation \eqref{eq:I_x} when the spatial distribution of optical 
depth along the LOS consists of some fraction $(1-C_\nu)$ of purely optically thin gas with $\tau_\nu = 0$, 
with the remaining gas optically thick with constant $\tau_\nu (= \tau_{\nu,r}$ for $I_r$), giving an 
unambiguous definition of the LOS covering fraction, $C_\nu$.  The PPC model reveals that any 
differences between the profiles $I_b$ and $I_r^2$ are due to partial covering ($C_\nu < 1$), and 
therefore doublet lines encode information about the distribution and optical depth of clouds.  
As we will show, they can also be used to infer the presence of cloud acceleration. 

We will invoke the PPC model to help interpret our results, so for consistency we should compare 
the properties of our simulations to this model's `doublet solution', i.e. its predictions for $C_\nu$ 
and $\tau_{\nu,r}$ when applied to our synthetic doublet lines, $I_r$ and $I_b$.
We must therefore calculate the covering fraction $C_\nu$ and a mean value of the optical depth, 
$\overline{\tau}_\nu$ from our simulations.  
To define either quantity, we require an operational definition of obscuration, 
i.e. a cutoff optical depth value, $\tau_{\rm{cut}}$, above which the background source is 
considered `covered'.  The fraction of sight lines with $\tau_\nu(x) > \tau_{\rm{cut}}$ defines $C_\nu$, 
while the average of $\tau_\nu(x)$ above this cutoff defines $\overline{\tau}_\nu$: 
 \begin{align}
 \begin{split}
 & C_\nu  =  \int_0^1 H[\tau_\nu(x) - \tau_{\rm{cut}} ] \,dx, \\
 &\overline{\tau}_\nu = C_\nu^{-1} \int_0^1 \tau_\nu(x) H[\tau_\nu(x)- \tau_{\rm{cut}}] \,dx.
\end{split}
\label{eq:Ctau_exact}
\end{align}
Here, $H(x)$ is the Heaviside step function.  
Note that these equations demand the profile $\tau_\nu(x)$ to be monotonically increasing.   
Provided the physical source is unresolved, so that an observer measures the integrated flux, 
$\tau_\nu(x)$ can always be sorted prior to integration. 

\begin{figure}
\includegraphics[width=0.5\textwidth]{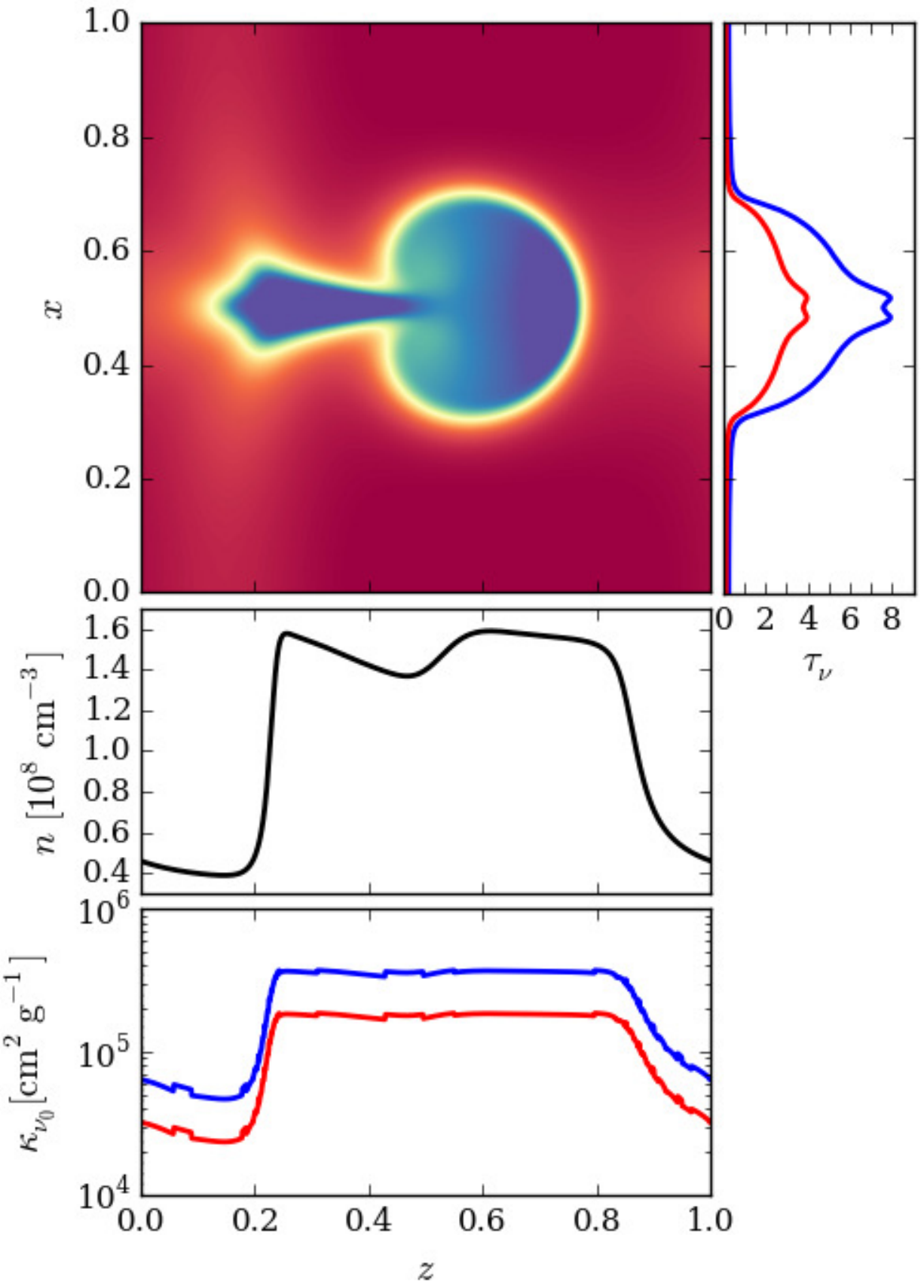}
\caption{
Illustrative calculation of the optical depth at line center, $\tau_{\nu_0}(x)$.  
The top row shows an image of the density field 
and the resulting spatial profile of $\tau_{\nu_0}(x)$.  
Both the $x$ and $z$ dimensions are measured in units of $L_x = 3.1\times10^{11}~\rm{cm}$.
We plot profiles of $\kappa_{\nu_0}(z)$ through the center of the domain ($x = 0.5$) for each component 
of the doublet in the bottom panel.  This quantity has been extracted from \textsc{xstar} using a lookup table 
based on the values of $(\xi,T)$, where $\xi$ is calculated from $n$, plotted in the middle panel.  
}
\end{figure}

\section{Results}
To illustrate our procedure for calculating simulated line doublets, in Figure~1 we display the relevant 
quantities that enter our calculations.   The top panel shows a 2D image of the number density 
along with an adjacent plot displaying the optical depth profiles of the 
two doublet lines along the $x$-axis.   
Here, $\tau_\nu(x)$ has been calculated at line center.   We will always take `line center' to mean at  
$v = -v_c$ in velocity space, corresponding to frequencies near $\nu_D = \nu_0(1+v_z/c)$ in frequency space.
The bottom panel shows sample profiles of $\kappa_{\nu_D}(z)$ for each transition along a horizontal slice 
through the center of the domain.  This profile, in turn, is calculated from the profiles of density and temperature, 
and density is shown in the middle panel.  Notice that for this snapshot, the density only varies by a factor of 4 
between warm and hot gas (as does temperature), while the line center opacity varies by a factor of about 60.  
This indicates that the ion fraction, $\eta_{\rm{ion}}$, is highly sensitive to the physical conditions of the gas.  

To compute the emergent flux at this line center frequency, we first attenuate a uniform background source 
according to these optical depth distributions.  A distant observer will measure this attenuated flux integrated 
over $x$.  To construct the profiles $I_r$ and $I_b$, we simply perform this integration and then repeat this 
procedure for all frequency bins.   

Note that because our domain size, $L_x = 3.1\times10^{11}~\rm{cm}$, is typically much less than the 
characteristic scale of an X-ray source 
($R_X \sim 10^{13}~\rm{cm}$ for a Seyfert galaxy with a black hole mass $10^7\textrm{M}_\odot$),
we are implicitly assuming a global covering fraction of clouds similar to the local covering fraction of warm gas
in our domain.  That is, in order for our calculations over a single domain to carry over to the full extent of the 
continuum source, we must envision an ensemble of $\sim( R_X/L_x)^2 \approx 10^3$ individual adjacent domains, 
each containing an evolving cloud.  Interpreted in terms of the cloud formation mechanism described in PW15, 
such a large number of clouds requires only the existence of an SED that can give rise to a thermally unstable 
region surrounding the continuum source.  Any number of perturbations can grow simply through a large change
in ionizing flux triggering cloud formation at a wavenumber, $k_{\rm{max}}$, corresponding to the maximum linear 
growth rate of the TI; $L_x = 2\pi/k_{\rm{max}}$ was assigned based on this characteristic length scale (see PW15).  
Due to the existence of this length scale, the distributions of resulting cloud sizes and densities are likely to be 
sharply peaked about some characteristic values that are only a function of the SED and photoionization parameter.
We therefore would not expect the basic results presented here to change if this `global' problem were to be 
directly modeled.  

\subsection{Synthetic absorption lines}
The results of our line profile calculations are presented
in Figures~2 and 3, where we show the time evolution of our two cloud configurations with and without a 
time-variable ionizing flux.  The leftmost panels in either figure are the same, as the variable flux is not 
applied before this time.  Each frame is separated by a half period ($t_X/2$) of the sinusoidal flux cycle, 
about 1.8 days (see \S{2.1}).  The velocity field of the gas is decomposed as 
$\vec{v}(x,z) = \delta v_x(x,z) \hat{x} +  [v_c + \delta v_z(x,z)] \hat{z}$, where $v_c$ is the 
velocity of the comoving frame that defines `line center' and $(\delta v_x, \delta v_z)$ denotes the local 
velocity field, whose directionality is depicted by the arrows, with the light grey to black color gradient 
representing the transition from negative to positive $\delta v_z$.  We display $v_c$ on the upper left 
corner of each panel in $\rm{km~s^{-1}}$, whereas the bracketed numbers in the bottom of each panel 
summarize the local LOS velocities: the left (right) number is the average over all velocities with 
$\delta v_z < 0$ ($\delta v_z > 0$).  The bottom rows of panels display the corresponding synthetic 
absorption lines for the \ion{O}{viii} doublet.

\begin{landscape}
\begin{figure}
\includegraphics[width=1.25\textwidth]{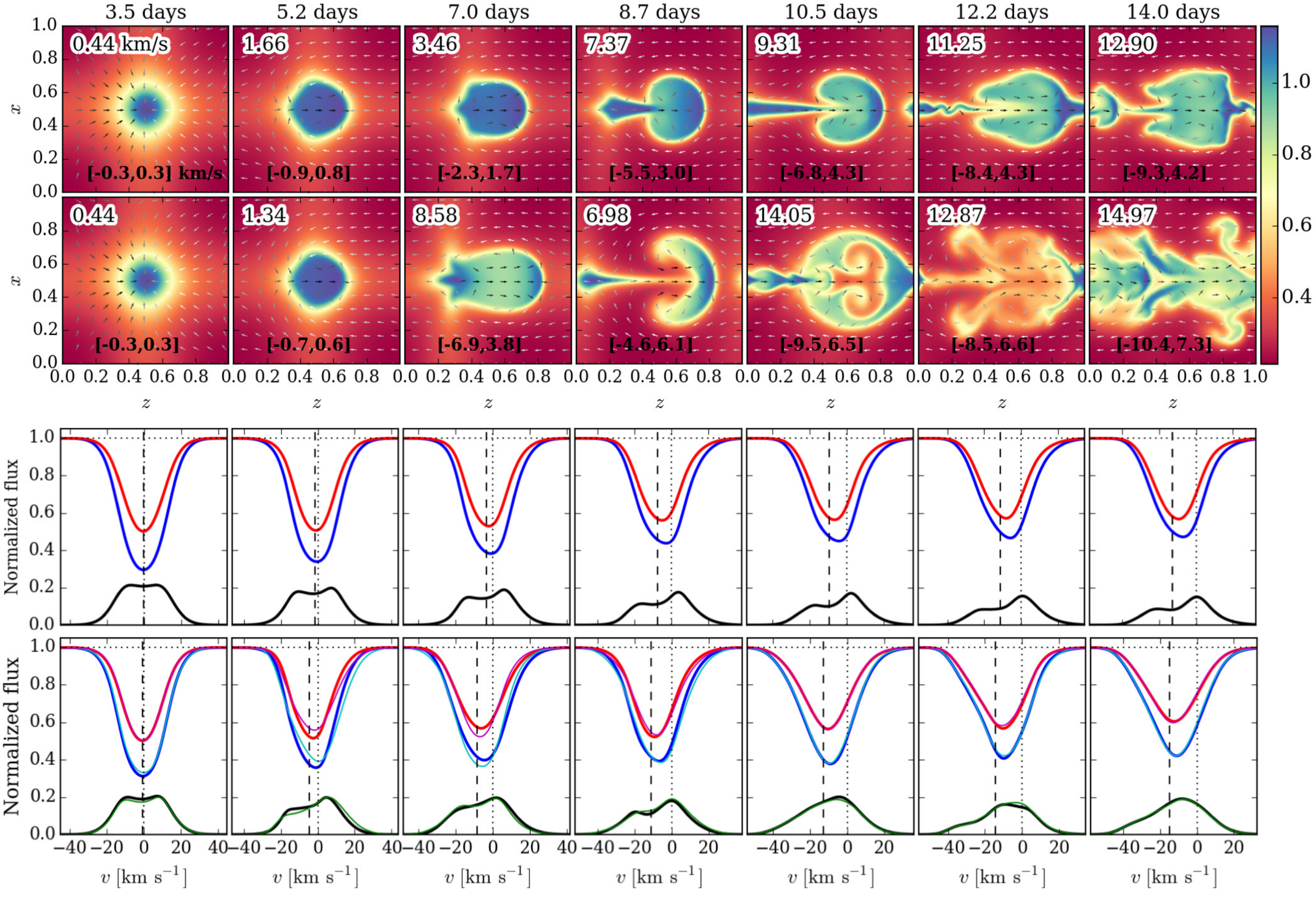}
\caption{Densiy maps (colorbar shows number density, $n$, in units of $10^{8}~\rm{cm}^{-3}$) 
and corresponding synthetic absorption line profiles for a constant flux run (first row of panels) 
and a variable flux run (second row of panels).  The early cloud formation stage is not shown.  
The source of radiation is plane parallel and emanates from the left (at $z = -\infty$), 
while the observer is to the right (at $z = \infty$).  The numbers in the top left of every density 
panel are the comoving velocities of the cloud, $v_c$, in $\rm{km~s^{-1}}$ and correspond to 
the vertical dashed lines in the bottom rows.  Arrows on the density maps depict the local velocity field, 
with the gradient in colors from light grey to black representing the transition from negative to positive 
local LOS velocities, $\delta v_z$.  The numbers in the brackets denote averages over all 
$\delta v_z <0$ and $\delta v_z >0$.  Blue and red lines denote $I_b$ and $I_r$, the stronger and 
weaker lines of the \ion{O}{viii} doublet, while the black lines show the difference $I_r - I_b$.  
To assess the line variability for the variable flux run, we overplot $I_b$, $I_r$, and $I_r - I_b$ using 
cyan, magenta, and green lines, respectively, for line profiles calculated at the later time $t + t_X/4$, 
which is either a high or low flux state; the density maps are always shown for a mean flux state.     
A partial covering analysis reveals that $I_r - I_b \approx C_\nu e^{-\tau_\nu}$, so dips in this profile 
can be caused by gas with higher optical depth, leading to the appearance of humps.  
The location of the dip tracks the faster moving, more opaque cloud material. 
}
\end{figure}
\end{landscape}

\begin{landscape}
\begin{figure}
\includegraphics[width=1.25\textwidth]{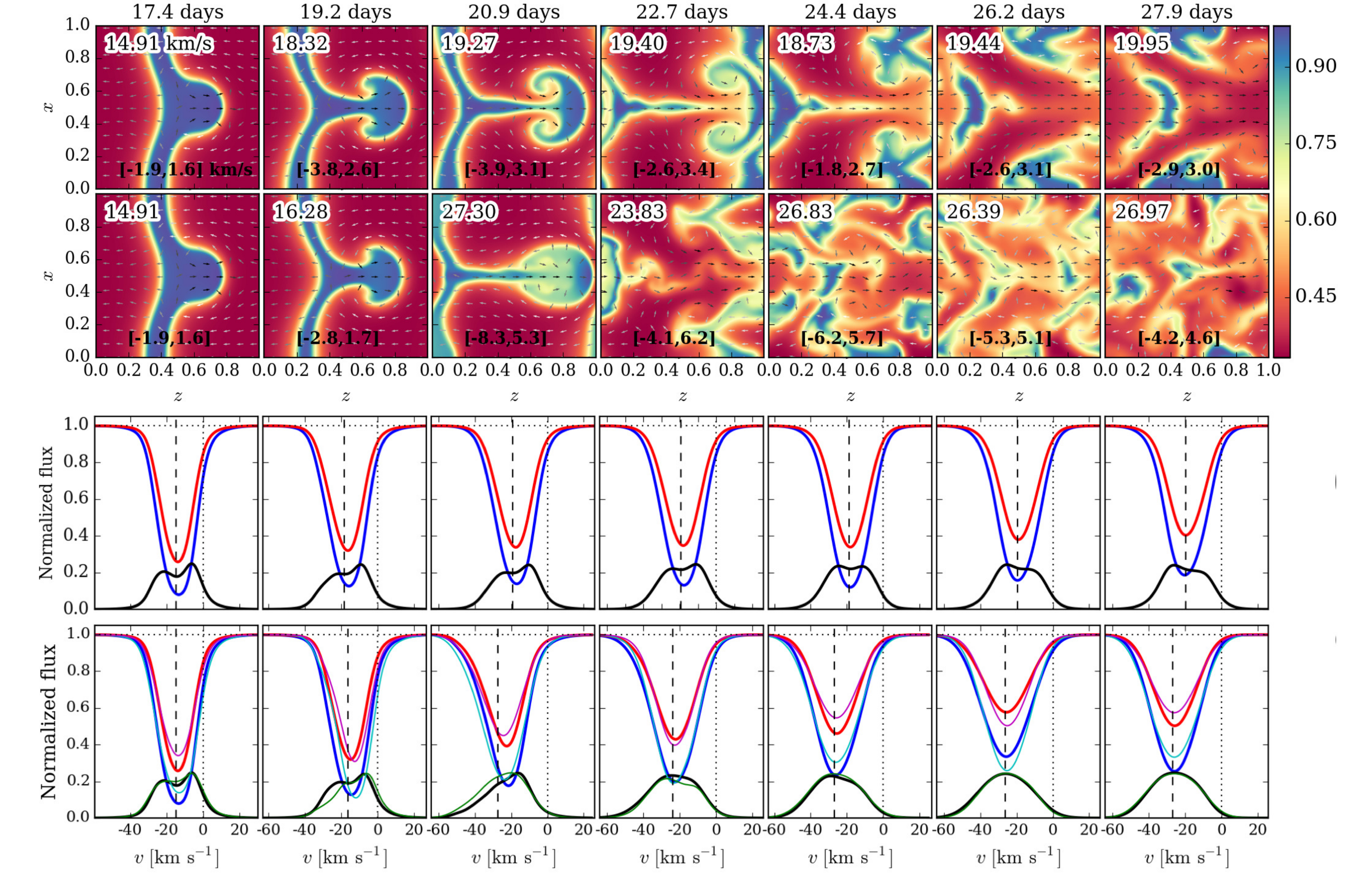}
\caption{Same is Figure~2 but for the slab geometry.  Compared to the round cloud case, 
there is less rapid acceleration and therefore the deeper absorption expected on the blue 
side side of line center is not actually visible on the line profiles.  However, the diagnostic 
$I_r - I_b$ still reveals the presence of large differences in the local velocity field.  Notice 
that in the bottom panels this signature of cloud acceleration gets wiped out as the cloud 
is disrupted since the velocities of the clumps become randomized. 
}
\end{figure}
\end{landscape}

\begin{figure*}
\includegraphics[width=\textwidth, resolution=80]{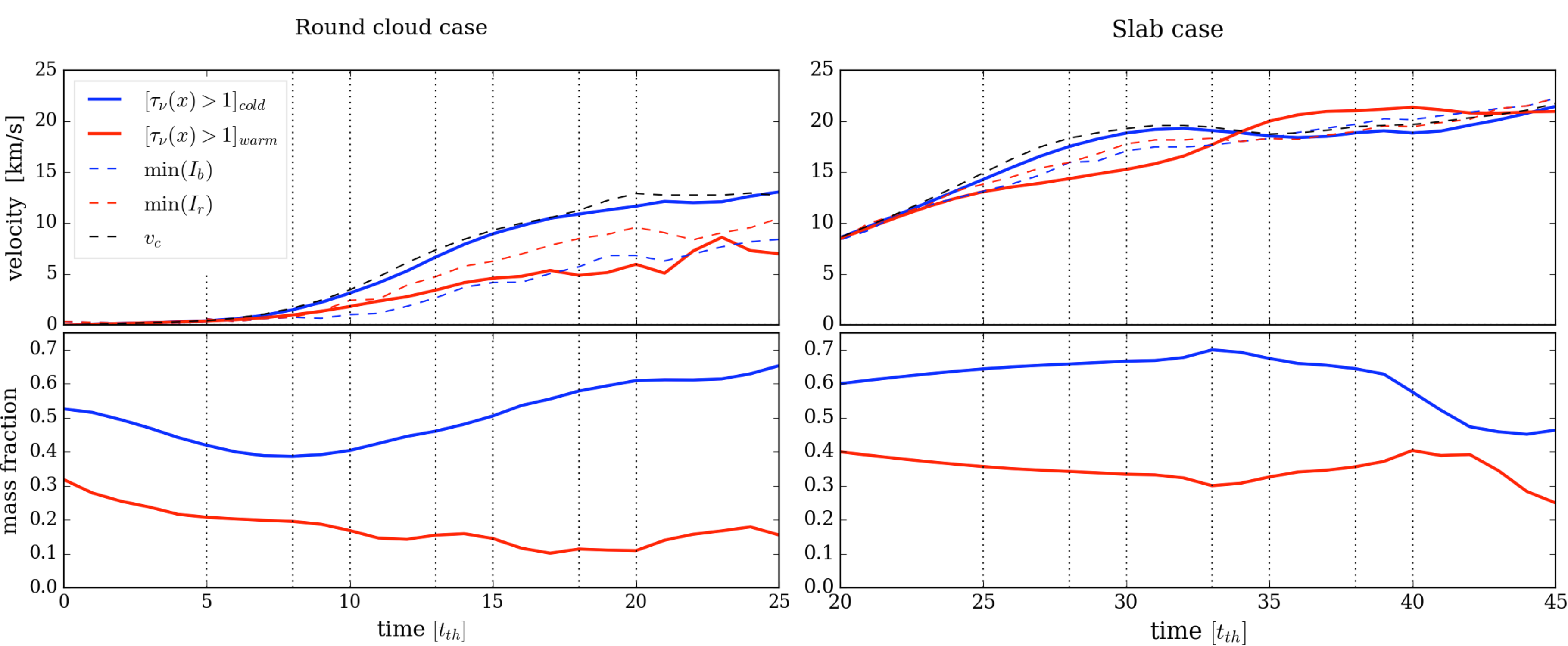}
\caption{
Temporal analysis linking cloud acceleration with the expected absorption properties.  The 
dotted vertical lines mark times corresponding to the snapshots in Figures~2 and 3.  
The thick solid red and blue curves denote the velocity (top panels) and mass fractions 
(bottom panels) averaged only along sight lines with high optical depth (i.e. only for $x$ 
with $\tau_\nu(x) > 1$).  The dashed red and blue lines trace the locations of line center 
(the minima of $I_r$ and $I_b$, respectively).  The dashed black line marks $v_c$, 
the average velocity all warm gas in the domain, and corresponds to the dashed vertical 
lines in Figures~2 and 3.  Since this curve coincides with the blue one, the dips in 
$I_r - I_b$ indeed trace the motion of cloud.  
In Figure~3 the center of these dips move to the right of $\min{I_r}$, which is shown more 
clearly here as the blue line pass beneath the red in the right panel, indicating that the cloud 
is moving slower than its surroundings on average.  
}
\end{figure*}

\begin{figure}
\includegraphics[width=0.5\textwidth]{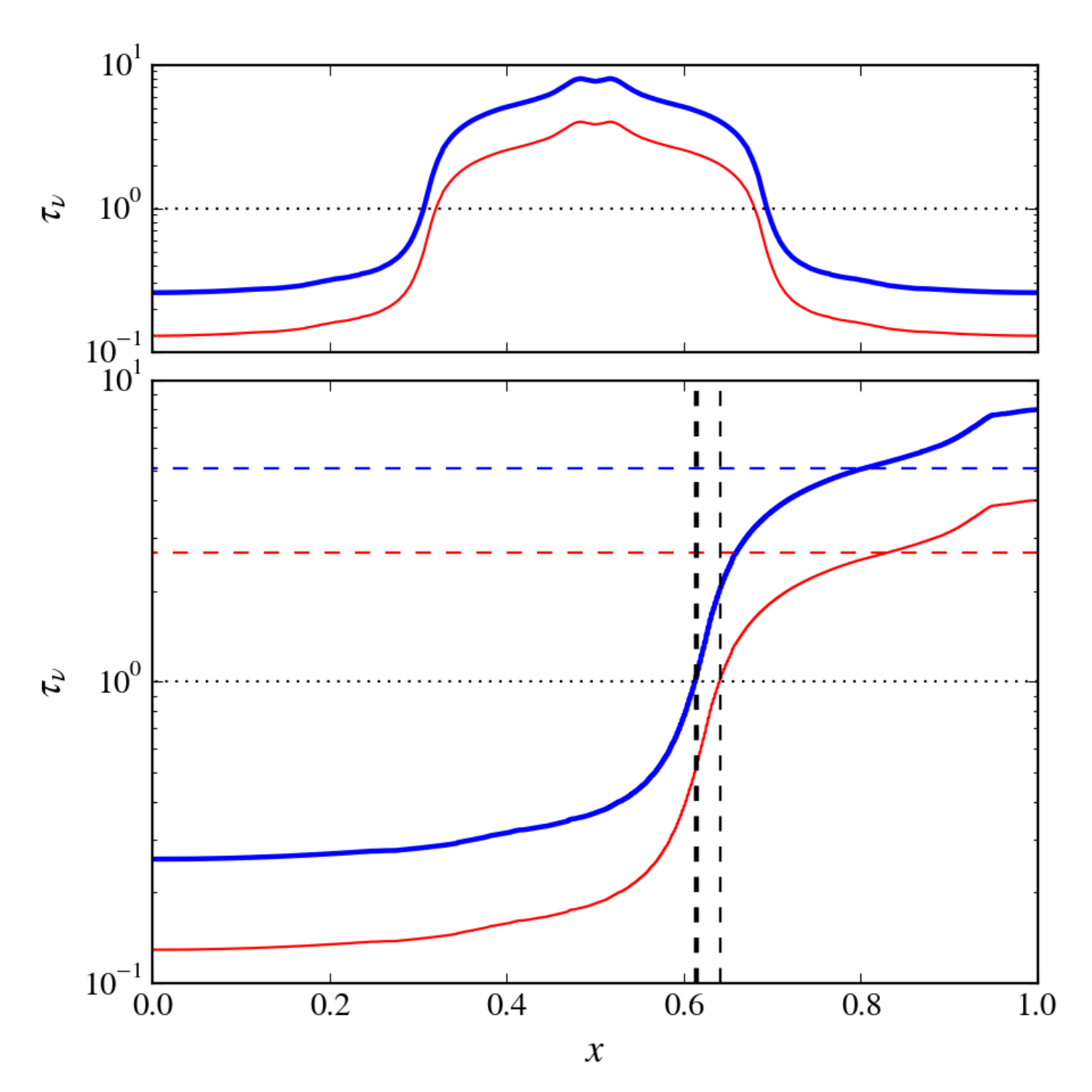}
\caption{
Method to calculate representative values of $\tau_\nu$ and $C_\nu$ in order to assess the 
PPC model.  The top panel reproduces the spatial optical depth profiles from Figure~1, but 
on a log-scale.  In the bottom panel, we sort these profiles in order of increasing optical depth.  
We then apply equation \eqref{eq:Ctau_exact}, taking $\tau_{\nu,\rm{cut}} = 1$, which is 
marked by the dotted line.  The blue and red horizontal lines show $\overline{\tau}_\nu$ for 
each transition, which is the average of $\tau_\nu(x)$ to the right of the thick and thin black 
vertical dashed lines, respectively.  The location of these vertical lines define $1 - C_\nu$ at 
line center.  Repeating this procedure at all frequencies, we arrive at the `exact' values plotted 
as a function of velocity in Figure~6.
}
\end{figure}

\begin{figure*}
\includegraphics[width=\textwidth]{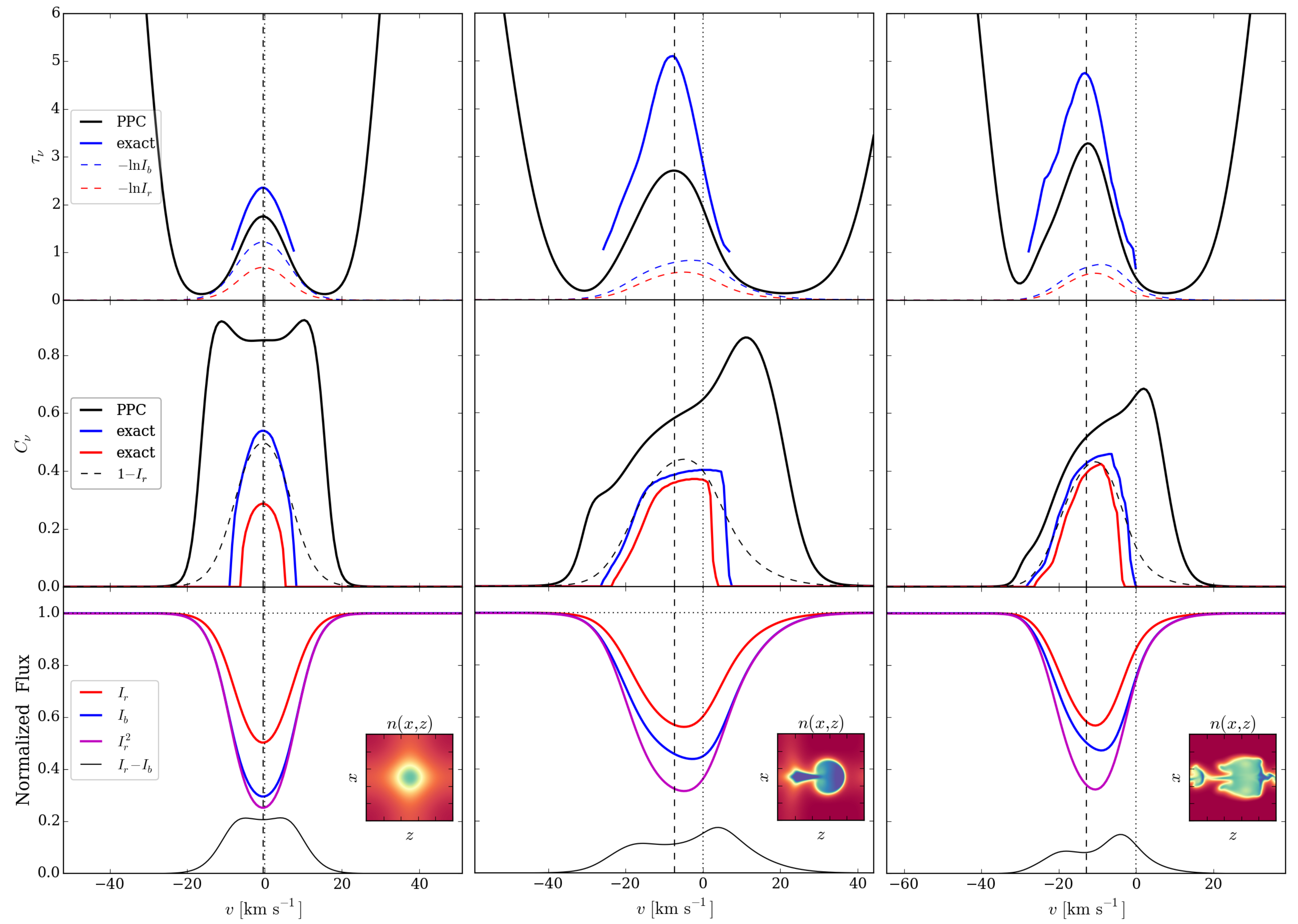}
\caption{
Comparison between the doublet solution of the PPC model (thick black curves) and the 
exact solutions for $\tau_\nu$ (top panel) and $C_\nu$ (middle panel).  Vertical dashed lines 
mark the comoving frame velocity, $v_c$.  The unphysical behavior of $\tau_{\nu,b} = 2\tau_{\nu,r}$ 
in the line wings is expected (see the discussion below equation \eqref{eq:dblet_soln}).  Notice 
that the PPC model's doublet solution underestimates the true optical depth profiles measured 
from our simulations, while it overestimates the covering fraction profiles.  
In the second and third columns, $C_\nu$ is almost the same for the red and blue doublet 
transitions, whereas they significantly differ in the first panel where $\tau_{\nu} < 3$ at line 
center.  Note that $C_\nu$ for the blue component is always close to its theoretical lower 
bound in the PPC model, $1 - I_r$ (dashed black curve).  Along with the line profiles of 
each doublet component (blue and red), the bottom panel displays $I_r^2$ (magenta) to 
assess the importance of partial covering.  
}
\end{figure*}

Our main result is that the absorption lines are overall smooth and nearly featureless, despite the 
dramatic changes 
in cloud morphology.  The reason for this is that the temperature distribution of 
the gas is the dominant factor in determining line shapes, as cloud acceleration and disruption does 
not generate enough velocity dispersion to significantly impact the line shapes.  
As indicated on the density images, the difference between the mean positive and negative LOS 
velocities increases with time, reaching $10-15~\rm{km~s^{-1}}$ in Figure~2 and 
$5-10~\rm{km~s^{-1}}$ in Figure~3.  These relative velocities do not exceed the thermal velocity 
for \ion{O}{viii} in the hot gas, which is about $20~\rm{km~s^{-1}}$. 
Thus, Doppler shifts due to the bulk velocities have a sub-dominant effect on the line widths. 

However, for the round cloud case the velocity dispersion due to cloud acceleration 
does exceed the thermal velocity for \ion{O}{viii} in the \emph{warm} gas, 
which is about $10~\rm{km~s^{-1}}$.  
Such systematic differences in velocity can viewed as velocity-space inhomogeneities in 
optical depth and visibly manifest as an asymmetry on the line profiles, as can be seen in 
Figure~2.  Beginning with the third panel, the blue line profiles develop an asymmetry, 
showing deeper absorption to the left of line center, which is the expected signature for 
cloud acceleration as explained in the introduction. 

Nevertheless, this asymmetry is admittedly subtle and 
moreover its presence is barely apparent for the slab geometry in Figure~3.  
Comparing the velocities from one frame to the next in Figure~3, 
it is clear that the bulk flow acceleration is substantially less compared to those in Figure~2, 
explaining the lack of a noticeable acceleration signature.  It is nevertheless possible to reveal 
its presence, as we now proceed to show.  

 \subsection{A spectral diagnostic for cloud acceleration}
Our analysis reveals that the difference $I_r - I_b$ leads to a profile whose shape is a 
sensitive diagnostic of cloud acceleration.  This quantity is plotted in black in the lower 
panels of Figures~2 and 3.  To assess the degree of variability, this quantity is also plotted 
in green in the bottommost row for a snapshot displaced in time a quarter cycle ($t_X/4$) 
ahead of that shown in the image, corresponding to a high or low flux state. 
Notice the overall trend among all cases: the profile for $I_r - I_b$ shows prominent asymmetric 
double humps when clouds are being accelerated, while the humps are small or altogether absent 
once the cloud is fully disrupted.  Importantly, these features are nearly as prominent in Figure~3 
as they are in Figure~2, despite the line profiles for the slab configuration being visibly symmetric 
and therefore lacking an obvious signature for cloud acceleration.

To explain these features, we have marked the values of $v_c$ with black dashed vertical lines, 
showing that they intersect the dips in $I_r - I_b$.  
Evidently these dips track the accelerating cloud and in turn lead to the appearance of humps.
Notice that these vertical lines are to the left of the locations of $\min(I_r)$ in Figure~2, 
implying that there is less residual flux at the blueshifted velocity $v_c$ compared to the same 
offset to the right of $\min(I_r)$, in agreement with our expectations.  

A qualitative understanding of the behavior of $I_r - I_b$ can be gained with the help of the 
PPC model.  From equation \eqref{eq:ppc_model}, we have  
\begin{align}
\begin{split}
I_r - I_b &= C_\nu e^{-\tau_{\nu,r}} ( 1 - e^{-\tau_{\nu,r}}),\\
& \approx
\left\{
\begin{array}{lr}
C_\nu e^{-\tau_{\nu,r}}  & \text{     (near line center);}\\
\tau_\nu C_\nu e^{-\tau_{\nu,r}}  & \text{(in the line wings).}\\
\end{array}
\right.
\end{split}
\label{eq:Inu_ppc}
\end{align}
We invariably find $C_\nu$ to have a flat or mildly increasing slope near line center 
(see \S{\ref{PCmodels}}). Hence, the only way for the profile of $I_r - I_b$ to lose monotonicity 
near line center is if there are velocity-space inhomogeneities in $\tau_\nu$.  In other words, 
on the basis of the PPC model, a fast moving clump should indeed carve out a blueshifted dip 
in the otherwise smooth profile of $I_r - I_b$.  

Note, however, that $v_c$ does not always become increasingly blueshifted with time; 
in Figure~3, the vertical dashed lines can lie to the right of the velocity corresponding to 
$\min(I_r)$, indicative of locally redshifted cloud motion.
To explain this occurrence and verify quantitatively that 
the dips in $I_r - I_b$ are always tracking the mean cloud motion, 
we extract gas properties only along sight lines for which $\tau_\nu(x) > 1$ 
(at a velocity corresponding to $\min(I_b)$) to exclude sight lines with only hot gas.  
This will eliminate $\sim1/3 - 1/2$ of all $x$-values in Figure~2 but few or no sight lines in 
Figure~3.  Within this subset of our domain, we then determine the average velocity and 
mass fraction of the warm ($T < T_{eq}$) and hot ($T > T_{eq}$) gas. 

In the top panel of Figure~4, we plot these quantities versus time as solid blue and red curves, 
along with $v_c$ (dashed black line) and the velocities corresponding to the locations of 
$\min(I_r)$ and $\min(I_b)$ (dashed red and blue lines, respectively).
For the round cloud case, we find a clear separation in velocities between warm and hot gas, 
with the average velocity of warm gas at all times exceeding those marking $\min(I_r)$ and 
$\min(I_b)$.  It also nicely traces the curve of $v_c$, which is a consistency check since $v_c$ 
was computed using a mass-weighted average over \emph{all} warm gas in \textsc{Athena}.  
For the slab case, on the other hand, there is only a marginal separation in velocities between 
warm and hot gas until $t = 34 t_{th}$, when the cloud material overall reverses direction.  
That is, it retains a net overall blueshift according to the total LOS velocity 
$v_z = v_c + \delta v_z$, but the local LOS velocity $\delta v_z$ becomes (on average) negative.  
This occurrence corresponds the last 3 panels in Figure~3, where it is clear that the clump in the 
center of the domain still has $\delta v_z > 0$, but the larger clumps at the top and bottom of the 
domain have $\delta v_z < 0$.  This sign change of the local LOS velocity is reflected in the profile 
of $I_r - I_b$, as the absorption dip tracing the warm gas shifts from the left side of $\min(I_r)$ to 
the right side.  

In the lower panels of Figure~4, the higher mass fraction of warm gas indicates that the shape of the 
line profiles should primarily be controlled by the dynamics of the cloud.  We can then ask why, 
since the warm gas is closely tracked by the speed $v_c$, are the line profiles not more sharply 
peaked around the dashed black vertical lines marking $v_c$?  The answer is that there is 
significant absorption in both the evaporating gas component (yellow colors in Figures~2 and 3) 
and the hot gas (red colors), so that the cloud is overall enshrouded in a background 
absorption profile.  We will return to this point below, as it implies that the hot gas has 
non-negligible optical depth, which is at odds with a basic assumption of the PPC model.   

\subsection{Partial covering analysis} 
\label{PCmodels}
We have already invoked the PPC model to qualitatively explain the behavior of the profile $I_r - I_b$.
Here we examine the predictions of this model more closely, by comparing 
our results against its `doublet solution', obtained by   
solving equation \eqref{eq:ppc_model} for $\tau_{\nu,r}$ and $C_\nu$:
 \begin{align}
 \begin{split}
& \tau_{\nu,r}  = -\ln I_r -\ln\left[\f{I_r - I_b}{I_r - I_r^2}\right] ,\\
& C_\nu  = \f{1}{1 + (I_b - I_r^2)/(1 - I_r)^2}.
\label{eq:dblet_soln}
\end{split}
\end{align}
These expressions are equivalent to those quoted in the literature (c.f. Arav, Korista, \& de Kool 2002),  
but the first now explicitly shows the `correction' to $-\ln I_r$, the prediction for $\tau_{\nu,r}$ 
in the case of no partial covering (when $I_b = I_r^2$).  It reveals the upper bound 
$I_b \leq I_r$, since obviously we must have $I_r > I_r^2$.  Meanwhile, the second 
expression immediately gives the lower bound $I_b \geq I_r^2$, because $C_\nu \leq 1$ 
by definition, with $C_\nu =1$ corresponding to $I_b = I_r^2$ as required. 
Thus, the behavior of these expressions as a function of $I_b$ and $I_r$ is evident:
the larger the difference between $I_b$ and $I_r^2$, the smaller the covering fraction 
and the larger the correction for the `true' optical depth of the absorber.  

However, whenever $I_b \rightarrow I_r$ the correction term to the optical depth will clearly blow up, 
and the profile for $\tau_{\nu,r}$ must therefore be truncated in the line wings when $I_r - I_b$ 
becomes very small.  Only near line center is this behavior physical, for in the limiting case of 
doublets with matching residual intensities ($I_b = I_r$) and $\tau_{\nu,r} = \infty$, we correctly 
recover $C_\nu = 1 - I_r$, which is a lower bound on the profile for $C_\nu$.  
In practice, application of the PPC model has typically been limited to the core of the line due to a
similar anomalous effect causing unphysical values for $C_\nu$ upon attempting to deconvolve the 
line profiles with the instrument's line spread function (e.g., Ganguly et al. 1999; Hall et al. 2003).

In the PPC model, the spatial optical depth profile is a step function at each frequency, 
which entails the following assumptions:  
(i.) the hot gas has zero optical depth;  
(ii.) the warm gas has
a single optical depth value (equal to $2\tau_{\nu,r}$ in the case of the stronger line of the doublet); 
and 
(iii.) the covering fractions of either line in the doublet are the same.  
We assess these assumptions at line center in Figure~5, by taking the spatial optical depth profile 
from Figure~1 as a case in point.  In the top panel of Figure~5, we plot $\tau_\nu(x)$ on a log-scale, 
showing that its minimum value is about 0.25 for the blue component, which is hardly negligible.  
In the bottom panel, we sort $\tau_\nu(x)$ to make it a monotonically increasing profile.  
We see that it indeed can be crudely approximated as a step function, with $\tau_{\nu,r}$ 
being represented by $\overline{\tau}_\nu$ (marked with a red dashed line), 
computed using equation \eqref{eq:Ctau_exact}.   
Additionally, the separation between the two vertical dashed lines measures the difference in 
covering fractions of either component (see the caption for details).  Since it is relatively small, 
we conclude that assumption (i) will be the most severe upon comparing the PPC model's 
doublet solution with our simulations. 
This comparison is carried out in the top two rows of Figure~6 for times corresponding to the first, 
middle, and last panels in Figure~2.  We indeed find that the predictions of the PPC model, 
the black solid profiles, do not agree well with the exact values measured from our simulations.  
However, we confirm in Appendix A that the agreement can be improved by accounting for the 
fact that the hot gas has a non-negligible optical depth.

\subsubsection{Effects of partial covering}
Notice from Figure~6 that the peak of the optical depth profile does not coincide with 
$\max{(C_\nu)}$ or $\min(I_b)$ in the second and third panels.  This is a partial covering effect: 
if we were to only calculate line profiles for sight lines in the middle half of the domain, this would 
not be the case.  In other words, the line profile due to the warm gas alone is almost completely 
saturated, so the residual flux around $v_c$ is coming entirely from the flux transmitted through 
the hot gas.  

The cloud is still in the process of forming in the left panel, and therefore the warm and hot gas 
only differ in density and temperature by about a factor of 2.  
This is why $I_b$ is approximately equal to $I_r^2$ in the bottom left panel, as the effects of 
partial covering are minimal in the absence of significant optical depth contrasts.   

Another partial covering effect is the tail appearing on the blue side of the $C_\nu$ profiles in the 
middle and right panels.  Since fast moving clumps tend to have higher $C_\nu$, this is in agreement 
with expectations.  We note that the PPC model's doublet solution, despite grossly overestimating 
the red side of $C_\nu$, correctly captures this overall feature on the blue side.  

\subsection{Line profile variability}
\label{sec:variability}
The blue and red line profiles in the last row of Figures~2 and 3 correspond to times in which the 
sine function is zero in equation \eqref{eq:Fion}.  Overplotted with these are line profiles corresponding 
to the nearest high or low flux state (when $\sin{(2\pi t/t_X)} = \pm 1$) to show the effects of variability 
in response to changes in the ionizing flux.   As shown by WP16, there are significant variations in 
temperature and density during these states, yet this evidently only amounts to a moderate level of flux 
variability.  Comparing the strengths of the lines across each row, we see that more variability is caused 
by cloud disruption than by the response of the gas to the 20\% changes in $\Delta F_X$.  
  
Indeed, comparing the cases without variability in the ionizing flux, 
we see that the time-dependent cloud dynamics alone causes significant absorption line variability.
Furthermore, in an environment prone to TI, a large change in $\Delta F_X$ (i.e. $A_X \gtrsim 1$ in equation \eqref{eq:Fion}) 
can trigger cloud formation by displacing 
$\xi$ to unstable regions on the radiative equilibrium curve (PW15), and this original deepening of the line 
profile (not shown here) can also be a source of variability. 

\section{Discussion}
We have combined 2D hydrodynamic simulations of the evolution of irradiated, thermally unstable gas in AGN 
with a dense grid of \textsc{xstar} models in order to calculate the detailed photoionization structure of the gas.  
This type of calculation represents the state of the art, as several research groups have recently developed 
interfaces between a hydrodynamic code and either \textsc{xstar} (Kinch et al. 2016) or \textsc{Cloudy} 
(Salz et al. 2015; Ram\'irez-Velasquez et al. 2016) in order to self-consistently account for radiation source terms 
under the assumption that photoionization equlibrium is established on time-scales short compared to all relevant 
dynamical time-scales (recall \S{2.2}).  Of these efforts, only Kinch et al. (2016) have also performed 2D simulations.  
Our approach here is not fully self-consistent, even within this approximation, as the heating and cooling rates and 
radiation force used in our hydrodynamic simulations are based on analytic fits to earlier photoionization calculations 
that assumed the same SED (see PW15 for details).  Here, \textsc{xstar} output is not being coupled to the 
hydrodynamics, as that first requires a non-trivial calculation of the force multiplier resulting from a given SED, 
work which is underway in our group (Dannen, Proga, \& Kallman 2016, in preparation).  
While we have yet to explore how different 
SEDs can affect the acceleration and dynamics of clouds, 
they are unlikely to change the qualitative outcome of our local cloud simulations, 
which show that the flow becomes turbulent but remains clumpy.

Several global models of the broad line region (BLR) exist in which cloud formation is attributed to TI.  
For example, Wang et al. (2012) considered a global model for episodic BLRs in which TI is 
ultimately responsible for the cloud production, with the origin of the gas being attributed to a Compton heated wind 
arising from the irradiated `skin' of a self-gravitating accretion disc.  Recently, Begelman \& Silk (2016) proposed the idea 
of magnetically elevated accretion discs and showed that TI should operate in the uppermost layers of the disc, 
thereby providing a different candidate scenario for a BLR.  Note, however, that TI is but one mechanism to produce a 
multiphase flow.  Others that have been considered in the context of AGNs include 
clouds uplifted from the surface of an accretion disc by the ram pressure of a centrifugally driven MHD wind 
(e.g., Emmering et al. 1992; Kartje, K\'{o}nigl, \& Elitzur 1999; Everett, K\'{o}nigl, \& Kartje 2001), 
transient density enhancements produced by turbulence (e.g., Bottorff \& Ferland 2001) in a radiatively driven wind, 
and a failed dusty wind (Czerny \& Hryniewicz 2011; Czerny et al. 2015).

The absorption line profiles calculated herein can be expected to be representative of those produced by other 
multiphase flows, provided the clumpy gas is only moderately optically thick ($\tau_\nu \lesssim 5$) and subsonic 
with respect to its surroundings.  Moreover, because these are local hydrodynamical simulations and the governing 
equations are Galilean invariant, our results apply equally well to clumps embedded in a highly supersonic outflow.  

An important result from this work is that a complicated velocity field is not accompanied by a complicated line profile.  
Quite the contrary, the most chaotic flow regime produces the most symmetric line profiles: in the final two panels of 
the slab case with a variable flux (see Figure~3), the profile $I_r - I_b$ is single peaked and featureless.  This is simply 
because the velocities of the clumps have been randomized and therefore 
(i.) any local acceleration signature has been lost and 
(ii.) the width of the absorption lines are set by thermal broadening alone because the local velocity field is subsonic.  
This result is consistent with several studies that have invoked `microturbulence' to broaden line profiles, as the level 
of microturbulence required is supersonic 
(e.g., Horne 1995; Bottorff et al. 2000; Bottorff \& Ferland 2002; Baldwin et al. 2004; Kraemer et al. 2007).  
Since it is unclear how supersonic motions can arise and persist (e.g., Kraemer et al. 2012), 
it is more likely that broad absorption lines are the result of ion opacity being spread over a wide range of
bulk velocities that are naturally present in a large scale outflow.  

In addition to line profiles maintaining their smooth shapes before and after the onset of cloud disruption, 
it is important to point out that they mostly retain their strengths also, aside from a small amount of variability 
(recall \S{\ref{sec:variability}}).  Evidently, the existence of a morphologically in tact cloud is not important; 
preservation of the mass fraction (or equivalently the filling factor) of warm gas is all that really matters, 
for then the overall ionization state will be unaffected.  
In part, this result is due to the fact that the hot gas in our simulations has a non-negligible optical depth, 
as revealed in \S{3.3}, so that significant absorption still takes place in the absence of cloud material.   
Thus, the maintenance of line strengths may to some degree be a limitation of our simulations, 
as the hot gas is only about 5 times hotter than the warm gas and is therefore still cool enough to host \ion{O}{viii} ions.  
Nevertheless, the relatively high optical depth of the inter-cloud gas in these simulations may mimic the presence of a 
much larger column of very optically thin gas.  Moreover, it is quite possible that cloud formation sites for warm absorbers 
indeed have low optical depth contrasts since resonance line opacities in the X-ray are typically much smaller than those 
in the optical or UV (e.g., Kallman 2010). 

UV lines in BAL outflows, on the other hand, are more likely to be very optically thick if the absorption sites are 
produced by clouds (e.g., Arav \& Li 1994).  More work is required to produce synthetic line profiles for the common 
UV doublet lines.  This requires two advances: (i.) much higher resolution simulations to resolve the high temperature 
and density contrasts.  This will permit the modeling of a clumpy medium in which the hot gas is hot enough to be 
very optically thin and the warm gas cool enough to host UV lines. (ii.) solving the equations of radiation hydrodynamics 
using the variable Eddington tensor closure (e.g., Jiang et al. 2012; Proga et al. 2014), as optically thick clouds will 
cast shadows and this effect must be properly taken into account.

In this work we focused on a doublet line in the soft X-ray simply because \ion{O}{viii} $\rm{Ly\alpha}$ 
was found to be the strongest line produced at these temperatures and photoionization parameters.  
Note that to resolve this doublet in actual spectra, a resolution of $R = \lambda/\Delta \lambda > 3500$ is required; 
this is beyond the capabilities of the X-ray Integral Field Unit being designed for the Athena+ mission 
(which has $R \lesssim 2400$ in the soft X-ray band; Barret et al. 2016), but it may be within the reach of Arcus 
(Smith et al. 2016) or Athena+'s high resolution X-ray grating spectrometer mission.
Without resolving the doublet components, the only detectable cloud signature would be an asymmetric line profile: 
prior to complete cloud disruption, cloud acceleration will produce deeper absorption on the blue side of line center, 
as in Figure~2.

We can speculate that the common UV doublet lines may present even clearer evidence for cloud acceleration 
using the $I_r - I_b$ diagnostic.  According to the PPC model, $I_r - I_b \approx C_\nu e^{-\tau_\nu}$ near line center, 
and since $\tau_\nu$ can be very large, the dips in this profile should be very prominent.  It is unclear, however, 
if outflows are sufficiently clumpy in the UV, as continuous disc wind models can explain broad absorption lines 
(e.g., Sim et al. 2010) and there is growing observational support for the line-driven disc wind scenario 
(e.g., Filiz Ak et al. 2014).  Interestingly, there has been a recent attempt to measure the bulk acceleration of outflows 
using a large sample of \ion{C}{iv} BAL troughs drawn from various Sloan Digital Sky Survey programs (Grier et al. 2016).  
We note that a local cloud acceleration signature is to be distinguished from such bulk acceleration signatures that arise 
on scales spanning the entire outflow. 
As we mention below, cloud acceleration can likely only be detected in narrow absorption lines.

\section{Conclusions}
Our simulations show that irrespective of the initial cloud geometry, cloud disruption generates vorticity and thus results 
in a complicated velocity field.  Our main conclusion is that this process of mixing has little influence on the line profiles 
and ultimately makes them more symmetric.  Were the warm gas to completely evaporate upon mixing, we would expect 
the line profile to significantly weaken, but it should still remain smooth since thermal broadening always dominates any 
bulk velocity broadening from a locally subsonic velocity field.
 
We confirmed our expectation that cloud acceleration imprints a deeper absorption signature on the blue side of line 
center before the cloud is completely disrupted.  Our most interesting finding is that a sensitive spectral diagnostic for 
assessing the presence of cloud acceleration can be obtained by examining the difference ($I_r - I_b$) of the 
absorption line profiles of a doublet line.  This quantity appears in the `doublet solution' of the commonly used PPC 
model (see equation \eqref{eq:dblet_soln}), and hence it should be possible to perform this analysis for common UV doublets.  

The detection of a cloud acceleration signature can probably only come from narrow lines that are not much broader than a 
thermal width, as the signature would not be present for BALs, which likely originate from smooth outflows.  
Repeated detections are needed to infer the presence of cloud acceleration, 
as the location of the dip in $I_r - I_b$ should become increasingly blueshifted with time.  
Behavior that would serve as evidence for cloud disruption is if the dip reverses its blueward trend or gradually disappears
after being monitored on a regular (perhaps daily) basis.  The detection of such a time-scale and the absence or presence 
of repeated episodes showing this effect would strongly constrain models for the AGN environment.

Finally, we showed that in our simulations, 
line strengths are more affected by cloud disruption than by changes in ionization in response to 
a (20\%) variation in the ionizing flux.  Absorption line variability that is correlated with fluctuations in the continuum light 
curve has recently been interpreted as evidence that changes in the ionizing continuum drives this variability 
(e.g., Filiz Ak et al. 2013; Wang et al. 2015; Goad et al. 2016), while the absence of such correlations is instead attributed 
to the transverse motion of clouds across the line of sight (e.g., Muzahid et al. 2016; Wildy et al. 2016).    
Cloud formation and subsequent disruption is also uncorrelated with $\Delta F_X$ and can occur entirely along the line of 
sight without clumps leaving the field of view.  
Thus, it is an alternative to the two most commonly invoked explanations for absorption line variability.  

\section*{Acknowledgments}
TW thanks Sergei Dyda and Jenny Greene for discussions related to this work, 
as well as Pat Hall for discussions and helpful comments on the manuscript.
We acknowledge support provided by the Chandra award TM6-17007X issued
by the Chandra X-ray Observatory center, which is operated by the
Smithsonian Astrophysical Observatory for and on behalf of NASA under
contract NAS8-39073.

\appendix
\section{A modified doublet solution}

\begin{figure}
\includegraphics[width=0.5\textwidth]{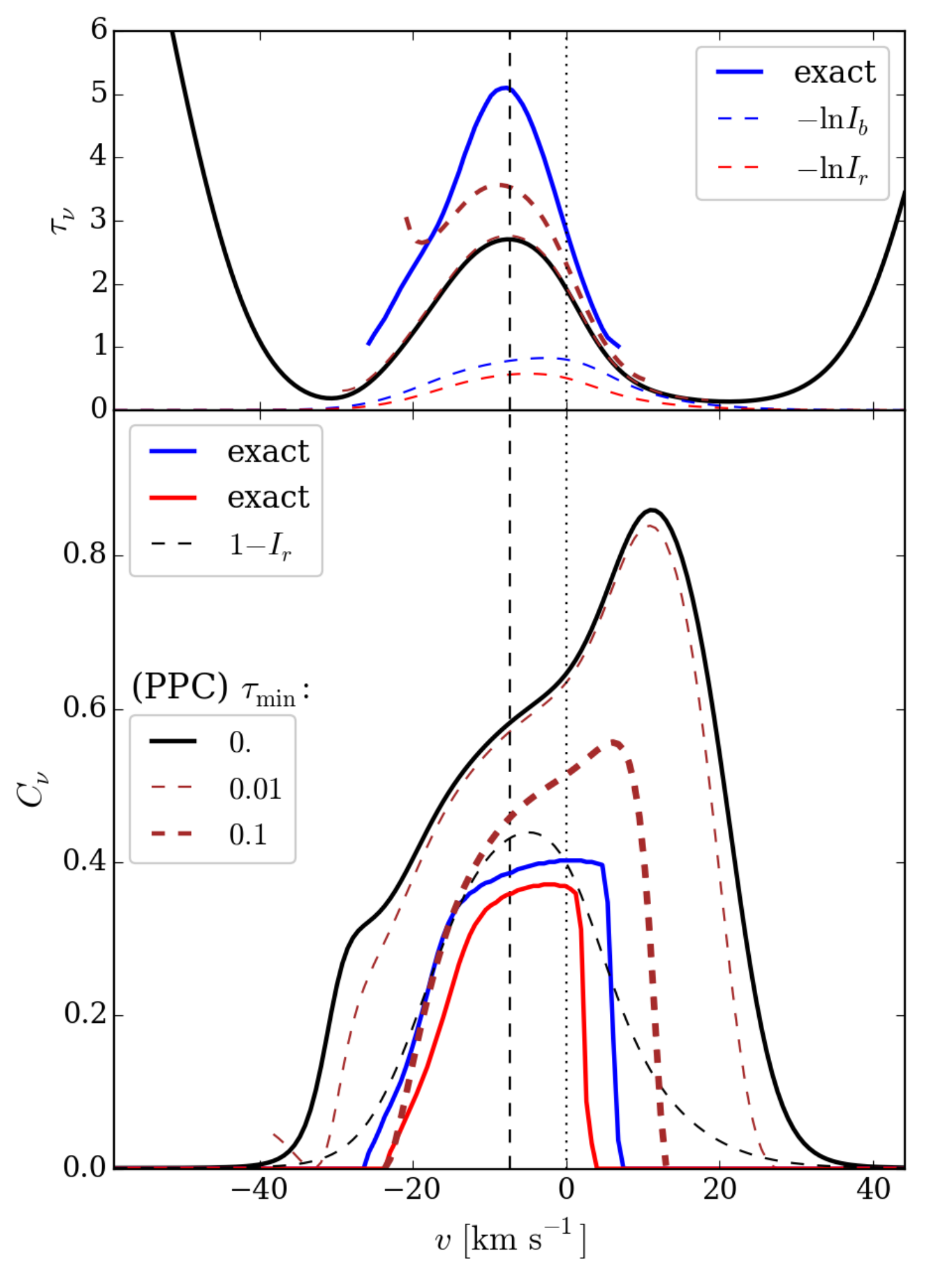}
\caption{
Same as the top two middle panels of Figure~6, but in addition showing two modified doublet solutions 
(dashed brown lines) that account for a nonzero value of $\tau_{\rm{min}}$.  These modified solutions 
also show unphysical behavior in the line wings, but we have truncated these curves for clarity.
This plot demonstrates that the discrepancy between the PPC model and the exact solutions becomes 
less with the modified doublet solution.   
}
\end{figure}

Here we modify the PPC model to account for the non-negligible optical depth in the hot gas, 
$\tau_{\nu,\rm{min}}$:
\begin{align}
\begin{split}
& I_r = (1-C_\nu) e^{-\tau_{\nu,\rm{min}}} + C_\nu\,e^{-\tau_{\nu,r}},\\
& I_b = (1-C_\nu) e^{-2\,\tau_{\nu,\rm{min}}} + C_\nu\,e^{-2\,\tau_{\nu,r}}.
\end{split}
\label{eq:mod_dblet_soln}
\end{align}
This model still entails that a given sight line passes through either hot gas or warm gas, but not both.
Dropping the frequency dependence of $\tau_{\nu,\rm{min}}$ and treating it as a constant free parameter, 
the modified doublet solution becomes
 \begin{align}
 \begin{split}
& \tau_{\nu,r}  = -\ln I_r -\ln\left[\f{I_r\, e^{-\tau_{\rm{min}}} - I_b}{I_r\, e^{-\tau_{\rm{min}}} - I_r^2}\right] ,\\
& C_\nu  = \f{1}{1 + (I_b - I_r^2)/(e^{-\tau_{\rm{min}}} - I_r)^2}.
\end{split}
\label{eq:mPPCsoln}
\end{align}
The brown dashed lines in Figure~A1 show how the doublet solution changes as 
$\tau_{\rm{min}}$ is increased from $0.01$ (thin brown line) to $0.1$ (thick brown line).  
For $\tau_{\rm{min}} = 0.1$, the discrepancy between the doublet solution and the profiles for 
$\tau_\nu$ and $C_\nu$ calculated from our simulations is significantly reduced.   
The agreement is not expected to become excellent since $\tau_{\rm{min}}$ should be frequency 
dependent instead of constant.  

The modified solutions are truncated at frequencies on either side of line center where 
$e^{-\tau_{\rm{min}}} = I_r$, since by equation \eqref{eq:mPPCsoln}, $C_\nu$ vanishes at these points.  
Notice that $C_\nu$ for the modified solutions with $\tau_{\rm{min}} =0.1$ can lie beneath $1-I_r$; 
this lower bound holds only for the original doublet solution.  
Also, the correction term for the optical depth now demands $I_b \leq I_r\,e^{-\tau_{\rm{min}}}$, 
and this should be interpreted as placing an upper bound on $\tau_{\rm{min}}$, namely $\tau_{\rm{min}} \leq -\ln(I_b/I_r)$.
Finally, by subtracting the expressions in equation \eqref{eq:mod_dblet_soln}, 
an extra term appears that will be negligible for $\tau_{\rm{min}} \lesssim 0.1$, 
so as to retain self-consistency in our application of the PPC model to qualitatively understand the behavior of $I_r - I_b$.  
\end{document}